\documentclass[twocolumn,pre]{revtex4}

\usepackage{dcolumn}
\usepackage{amsmath}

\usepackage{graphicx}

\begin{document}

\renewcommand{\d}{d}
\newcommand{\Ord}{\mathrm{O}}
\newcommand{\e}{\mathrm{e}}
\newcommand{\ii}{\mathrm{i}}
\newcommand{\half}{\mbox{$\frac12$}}
\newcommand{\set}[1]{\lbrace#1\rbrace}
\newcommand{\av}[1]{\left\langle#1\right\rangle}
\newcommand{\eref}[1]{(\ref{#1})}
\newcommand{\etal}{{\it{}et~al.}}
\newcommand{\defn}{\textit}
\newcommand{\mat}{\mathbf}
\renewcommand{\vec}{\mathbf}

\newlength{\figurewidth}
\setlength{\figurewidth}{0.95\columnwidth}
\setlength{\parskip}{0pt}
\setlength{\tabcolsep}{6pt}
\setlength{\arraycolsep}{2pt}

\title{Analysis of weighted networks}
\author{M. E. J. Newman}
\affiliation{Department of Physics and Center for the Study of Complex
Systems,\\
University of Michigan, Ann Arbor, MI 48109}
\affiliation{Santa Fe Institute, 1399 Hyde Park Road, Santa Fe, NM 87501}

\begin{abstract}
The connections in many networks are not merely binary entities, either
present or not, but have associated weights that record their strengths
relative to one another.  Recent studies of networks have, by and large,
steered clear of such weighted networks, which are often perceived as being
harder to analyze than their unweighted counterparts.  Here we point out
that weighted networks can in many cases be analyzed using a simple mapping
from a weighted network to an unweighted multigraph, allowing us to apply
standard techniques for unweighted graphs to weighted ones as well.  We
give a number of examples of the method, including an algorithm for
detecting community structure in weighted networks and a new and simple
proof of the max-flow/min-cut theorem.
\end{abstract}
\maketitle

\section{Introduction}
Many systems can usefully be represented as networks or
graphs---collections of vertices joined in pairs by edges.  Examples
include the Internet and the world wide web, citation networks, social
networks, and biological and biochemical networks of various kinds.
Although an old and well established branch of study in mathematics and
sociology, research on networks has in recent years attracted significant
attention from members of the physics community as well, who have
successfully applied a variety of physical ideas to the analysis and
modeling of these systems~\cite{Strogatz01,AB02,DM02,Newman03d}.

Most of the networks that have been studied in the physics literature have
been binary in nature; that is, the edges between vertices are either
present or not.  Such networks can be represented by $(0,1)$ or binary
matrices.  A network with $n$ vertices is represented by an $n\times n$
adjacency matrix~$\mat{A}$ with elements
\begin{equation}
A_{ij} = \biggl\lbrace\begin{array}{ll}
           1 & \qquad\mbox{if $i$ and $j$ are connected,}\\
           0 & \qquad\mbox{otherwise.}
         \end{array}
\end{equation}
However, as has long been appreciated, many networks are intrinsically
weighted, their edges having differing strengths.  In a social network
there may be stronger or weaker social ties between individuals.  In a
metabolic network there may be more or less flux along particular reaction
pathways.  In a food web there may be more or less energy or carbon flow
between predator-prey pairs.  Edge weights in networks have, with some
exceptions~\cite{Newman01c,Yook01,NR02,BBPV04,BBV04}, received relatively
little attention in the physics literature for the excellent reason that in
any field one is well advised to look at the simple cases first (unweighted
networks) before moving on to more complex ones (weighted networks).  On
the other hand, there are many cases where edge weights are known for
networks, and to ignore them is to throw out a lot of data that, in theory
at least, could help us to understand these systems better.

In this paper, we highlight a simple but useful technique that allows us to
say many things about weighted networks without deviating far from the
familiar territory of unweighted ones.  By mapping weighted networks onto
multigraphs, we point out, many of the standard techniques that have been
developed to study unweighted networks can be carried over with little or
no modification to the weighted case.  We present a number of examples of
applications to well known network problems.

\section{Weighted networks and multigraphs}
A weighted network can be represented mathematically by an adjacency matrix
with entries that are not simply zero or one, but are equal instead to the
weights on the edges:
\begin{equation}
A_{ij} = \mbox{weight of connection from $i$ to $j$}.
\end{equation}
For example:
\begin{equation}
\setlength{\arraycolsep}{4pt}
\raise-9.5ex\hbox{\includegraphics[width=2.5cm]{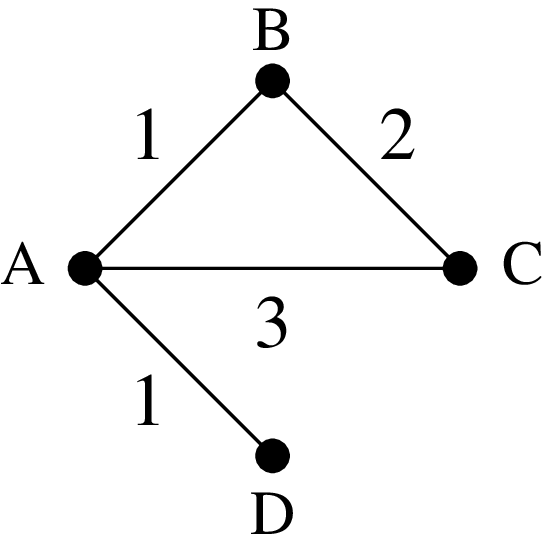}}\quad
  \raise-1.8ex\hbox{$\equiv$}\quad
    \begin{matrix}
      \hbox to 4.5em{\vphantom{\Big|}\footnotesize A\hfil B\hfil C\hfil D}
      \\
      \begin{pmatrix}
        0 & 1 & 3 & 1 \\
        1 & 0 & 2 & 0 \\
        3 & 2 & 0 & 0 \\
        1 & 0 & 0 & 0
      \end{pmatrix}
      &
      \begin{matrix}
        \mbox{\footnotesize A} \\
        \mbox{\footnotesize B} \\
        \mbox{\footnotesize C} \\
        \mbox{\footnotesize D}
      \end{matrix}
    \end{matrix}
\label{example1}
\end{equation}
In this example the weights on the edges are all integers, and we will
focus on the integer case for the moment.  We will also assume throughout
this paper that all weights are non-negative.  Negative weights are
possible in some cases.  They are, for instance, used sometimes in
sociological studies of acquaintance networks to represent animosity
between individuals.  We will not treat this case here however.

Adjacency matrices with non-negative integer entries occur in another
situation as well, in networks with multiple edges between vertex pairs.
For example:
\begin{equation}
\setlength{\arraycolsep}{4pt}
\raise-9.5ex\hbox{\includegraphics[width=2.5cm]{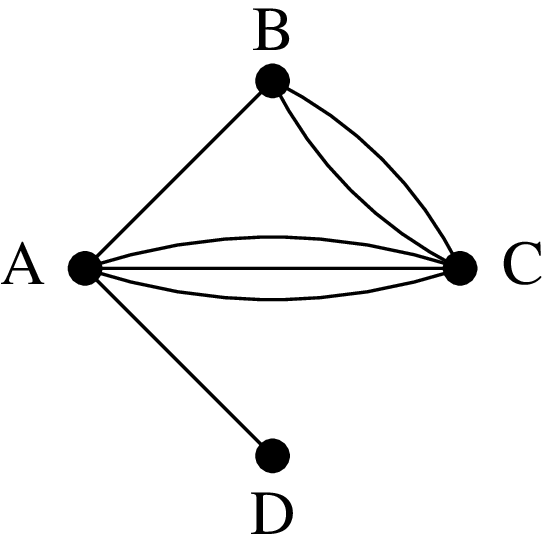}}\quad
  \raise-1.8ex\hbox{$\equiv$}\quad
    \begin{matrix}
      \hbox to 4.5em{\vphantom{\Big|}\footnotesize A\hfil B\hfil C\hfil D}
      \\
      \begin{pmatrix}
        0 & 1 & 3 & 1 \\
        1 & 0 & 2 & 0 \\
        3 & 2 & 0 & 0 \\
        1 & 0 & 0 & 0
      \end{pmatrix}
      &
      \begin{matrix}
        \mbox{\footnotesize A} \\
        \mbox{\footnotesize B} \\
        \mbox{\footnotesize C} \\
        \mbox{\footnotesize D}
      \end{matrix}
    \end{matrix}
\label{example2}
\end{equation}
Multiple edges are sometimes called \defn{multiedges} and networks or
graphs containing them \defn{multigraphs}, and we will use this
nomenclature here.

As we can see, the two networks \eref{example1} and~\eref{example2} have
the same adjacency matrix, and in many ways they behave the same.  For
example, if we are thinking of traffic flowing over the Internet (or even
traffic down a road), then the maximum traffic that can flow between two
vertices joined by two identical edges is the same as the maximum that can
flow between the same two vertices if they are joined by a single edge with
twice the capacity.  This suggests that we could obtain insight into the
behavior of weighted graphs very simply by mapping them onto unweighted
multigraphs.  That is, every edge of weight $n$ is replaced with $n$
parallel edges of weight 1 each, connecting the same vertices.  The
adjacency matrix of the graph remains unchanged and any techniques that can
normally be applied to unweighted graphs can now be applied to the
multigraph as well.

\subsection{Some simple examples}
Let us begin our demonstration of the principles above by giving a few
extremely simple examples of their use.  For our first, we ask what the
equivalent is of vertex degree in a weighted graph.  Recall that the degree
of a vertex is the number of edges attached to it.  We could use the same
definition for a weighted graph---simply count the number of edges attached
to a vertex regardless of their weight---but this, as we have said, ignores
much potentially useful information contained in the weights.  To the
extent that degree is a measure of the importance of a vertex in a network,
surely vertices with strong connections should be accorded more importance
than vertices with only weak connections?

If we apply our rule, mapping the weighted network to a multigraph, and
then calculate the degree as we would for a normal unweighted graph, we
immediately find that the degree~$k_i$ of a vertex~$i$ in a weighted
network is the sum of the weights of the edges attached to it:
\begin{equation}
k_i = \sum_j A_{ij}.
\label{degree}
\end{equation}
This certainly seems reasonable, and has indeed been proposed previously
using heuristic arguments~\cite{BBPV04}.  It also gives sensible results.
For instance, in a social network in which the weights on edges represent
the number of hours a person spends per week with each of their
acquaintances, their degree would be the total number of hours they
socialize per week---a very reasonable measure of social influence.  In the
case of traffic or current of some kind flowing around a network, with
weights representing the magnitude of the flow along the edges, the sum of
the flows on each of the edges attached to a vertex gives the total amount
of traffic flowing through the vertex.  In a road network for example the
degree of an intersection would just be proportional to the number of cars
passing through it.

As another example, consider eigenvector
centrality~\cite{Bonacich72a,Bonacich87,WF94}, a measure of centrality akin
to an extended form of degree centrality and closely related to
``PageRank'' and similar centrality measures used in web search
engines~\cite{BP98,Kleinberg99a}.  The eigenvector centrality $x_i$ of a
vertex in an unweighted network is defined to be proportional to the sum of
the centralities of the vertex's neighbors, so that a vertex can acquire
high centrality either by being connected to a lot of others (as with
simple degree centrality) or by being connected to others that themselves
are highly central.  We write
\begin{equation}
x_i = \lambda^{-1} \sum_j A_{ij} x_j,
\end{equation}
where $\lambda$ is some constant.  In matrix notation this becomes $\lambda
\vec{x} = \mat{A}\vec{x}$, so that $\vec{x}$ is an eigenvector of the
adjacency matrix.  By simple arguments one can show that one should take
the eigenvector corresponding to the leading eigenvalue~\cite{Friedkin91}.

By mapping to a multigraph, we can find the equivalent centrality measure
for weighted networks.  Network neighbors that are connected to a vertex
with twice the weight now contribute twice as much to the vertex's
eigenvector centrality.  As a result, we find that the correct
generalization of eigenvector centrality to a weighted network is, as we
would hope, still the leading eigenvector of the adjacency matrix, with the
elements of the matrix being equal to the edge weights, as before.  Such a
measure could be useful for example for ranking search results in a
citation network~\cite{Price65,Redner98}.  If a paper cites another many
times rather than just once, it could be taken as an indication of a closer
or stronger connection between the two papers.  Using such citation
frequencies as edge weights, our eigenvector centrality would then give
papers high scores either if they are cited by many others or if they are
cited with high weight by a few others.

Many authors have studied random walks on
networks~\cite{AKS03,Tadic03,NR04}.  What should be the appropriate
generalization of walks to weighted networks?  Mapping the network to a
multigraph and then performing an ordinary random walk on the resulting
unweighted network, we get a walk that traverses edges always in proportion
to their weight.  Thus at vertex~$i$ the walk chooses a step to vertex~$j$
with probability
\begin{equation}
P_{ij} = {A_{ij}\over\sum_j A_{ij}} = {A_{ij}\over k_i},
\label{rw}
\end{equation}
which is exactly the same rule we use for walks on unweighted graphs,
provided we generalize the definition of the degree~$k_i$ as in
Eq.~\eref{degree}.  Again this is an intuitively sensible result.  If we
have something flowing around a network, such as water through a network of
pipes, then Eq.~\eref{rw} is precisely the rule that would be followed by a
passive ``tracer'' molecule swept along by the water, provided that the
water is well mixed at each network node, so that we get a random walk
rather than some kind of correlated walk.

\subsection{The max-flow/min-cut theorem}
\label{maxflow}
The results above are all, in a sense, trivial, though it is satisfying
that our simple rule for understanding weighted networks leads us to them
naturally.  Now let us turn to some more substantive applications.  First,
we use our mapping to multigraphs to rederive a famous result in the theory
of networks, the max-flow/min-cut theorem.

The max-flow/min-cut theorem is a theorem about weighted networks.  It
states that, in a network where the weights represent the maximum allowed
flow of a fluid or other commodity along the edges, the following is true:
\begin{quote}
The maximum flow that can pass between any two vertices is equal to the
weight of the minimum edge cut set that separates the same two vertices.
\end{quote}
An \defn{edge cut set} is a set of edges whose removal from the graph will
disconnect the vertices in question.  A \defn{minimum edge cut set} is a
cut set of edges the sum of whose weights has the minimum possible value
for such a set.  The weight of the minimum cut set is called the
\defn{connectivity} of the vertices in question.

The equality of maximum flow and minimum cut set size has an important
practical consequence.  There are simple computer algorithms, such as
preflow-push algorithms~\cite{AMO93}, that can calculate maximum flows
quickly (in polynomial time), and the equivalence implied by the
max-flow/min-cut theorem means that we can use the same algorithms to
calculate sizes of minimum cut sets as well.  Maximum flow algorithms are
now the standard numerical technique for calculating sizes of cut sets.

Here we show that the max-flow/min-cut theorem can be deduced from a much
earlier and simpler theorem about unweighted networks, Menger's theorem.
Menger's theorem is often derived as a corollary of the max-flow/min-cut
theorem, but we show that the derivation can proceed in the opposite
direction as well.  This is interesting for two reasons.  First, it offers
a quite different proof of the max-flow/min-cut theorem from the usual one,
which is based on augmenting paths and residual graphs.  Second, it is
considerably harder to prove the max-flow/min-cut theorem from first
principles than it is Menger's theorem, so the method we describe offers a
more transparent demonstration of the max-flow/min-cut theorem than the
usual textbook presentations.

Menger's 1927 theorem states the following for an unweighted
network~\cite{Menger27}:
\begin{quote}
If there exists no cut set of fewer than $n$ edges between two vertices in
a graph, then there are at least $n$ edge-independent paths between the
same two vertices.
\end{quote}
Two paths through a network are said to be edge-independent if they share
none of the same edges~\footnote{In fact, Menger originally stated his
theorem for vertex cut sets and vertex-independent paths, but the extension
to edges is trivial and easily proved.}.  Many proofs of Menger's theorem
have been given---see, for instance, Ref.~\cite{West96}.

Given Menger's theorem, we first establish the truth of the
max-flow/min-cut theorem for unweighted networks as follows.  Consider the
maximum flow between two vertices $s$ and~$t$ in a network and suppose that
a minimum edge cut set between these vertices consists of $n$ edges.  The
removal of any edge in this cut set will reduce the flow by at most one
unit, since that is the maximum flow an edge can carry in an unweighted
network.  Thus if we remove all $n$ edges in the cut set one by one, we
remove at most $n$ units of flow.  But, since the cut set disconnects the
vertices $s$ and~$t$, this removal must stop all of the flow.  Hence the
entire flow is at most~$n$.

However, Menger's theorem tells us that if the minimum cut set has size~$n$
then there must be at least $n$ edge-independent paths between $s$ and~$t$.
Each of these paths can carry a single unit of flow from $s$ to~$t$ and
hence the network as a whole can carry at least~$n$ units between these two
vertices.

Thus the maximum flow between $s$ and~$t$ is simultaneously both at most
and at least~$n$, and hence it must in fact be exactly equal to~$n$: the
maximum flow is equal to the size of the minimum cut set in an unweighted
graph.  Note that this result applies just as well to graphs with
multiedges as to those with only single edges.

Now we extend this result to weighted graphs using the mapping between
weighted graphs and multigraphs.  If we take a network of pipes and replace
every pipe that can carry a maximum of $n$ units of flow by $n$ pipes that
can carry one unit each, then the maximum flow between any adjacent pair of
vertices is unchanged, and hence the maximum flow between any two vertices
in the network is also unchanged.

Now every minimum cut set on an unweighted multigraph includes either all
or none of the parallel edges between any adjacent pair of vertices; there
is no point cutting one such edge unless you cut all of the others as
well---you have to cut all of them to disconnect the vertices.  Thus, the
minimum cut set consists of sets of cuts of all the edges between certain
vertex pairs.  If we consider all such cut sets, minimal or not, and then
minimize over them, we will find the global minimum cut set.  However,
these cut sets are in a trivial one-to-one correspondence with, and have
the same weight as, the cut sets on the weighted graph, and hence the
minimum cut set on the weighted graph has the same weight as that on the
multigraph.

Thus both maximum flows and minimum cuts are numerically equal on
unweighted multigraphs and the corresponding weighted graphs, and hence
since the max-flow/min-cut theorem is true on unweighted graphs---including
multigraphs---it must also be true on the corresponding weighted graphs.

Finally, we extend the result to the case of non-integer weights.  To do
this we simply redefine what we mean by a unit of flow.  Let the size of
the unit of flow be~$r$.  Then a weighted edge with maximum flow~$nr$ for
$n$ integer transforms into $n$ edges of flow~$r$ each in the multigraph.
The proof goes through as before, and as we allow $r\to0$, all values of
the edge weights are allowed and hence the max-flow/min-cut theorem is
proved for all weighted networks.

This last trick, of changing the size of the units which we use to
transform weighted edges into unweighted multiedges, can be used for many
calculations or proofs for weighted graphs, and this relaxes the assumption
we made earlier that the weights in the graph are integers.  In this way,
essentially all the results presented in this paper can be extended to the
non-integer case also.

\subsection{Community structure in weighted networks}
We turn now to a quite different question about weighted networks, that of
community structure.  Many networks consist not of an undifferentiated mass
of linked vertices, but of distinct ``communities''---groups of vertices
within which the connections are dense but between which they are sparser.
This type of structure is seen especially in social networks, but also in
some biological and technological networks as well.  An interesting problem
that has attracted much attention in recent years is that of creating a
computer algorithm which, when fed the topology of a network, can extract
from it the communities in the network, if there are any.  The problem is
related to the problem of graph partitioning, which is well studied in
computer science, but algorithms for graph partitioning, such as the
Kernighan--Lin algorithm~\cite{KL70} or spectral
bisection~\cite{Fiedler73,PSL90} are not ideally suited to general network
analysis: typically they only divide networks in two, rather than into a
general number of communities, they provide no measure of how good the
division in question is, and in some cases they also require the user to
specify the sizes of the communities before they start.  In general they
also work only on unweighted networks.

Recently, Girvan and Newman~\cite{GN02} proposed an algorithm for community
structure discovery in unweighted networks that avoids these drawbacks and
appears to work well for many kinds of networks.  Since the publication of
that work, the author has been asked a number of times whether an
appropriate generalization of the algorithm exists for weighted networks.
Certainly the algorithm can be applied to such networks by simply ignoring
edge weights, but, as we have argued in this paper, to do so is to throw
away useful information contained in the weights, information that could
help us to make a more accurate determination of the communities.  In this
section we use the techniques discussed in this paper to derive an
appropriate generalization of the algorithm of Girvan and Newman to
weighted networks.

It is worth pointing out, before proceeding, that not all weights on
network edges are necessarily appropriate as input for determining
community structure.  Weights that indicate particularly close connections
or similarity between vertices can give useful information about
communities, but one can also put many other kinds of variables on edges,
and they certainly need not be indicators of proximity or similarity.  For
example, Barrat~\etal~\cite{BBPV04} have studied the network of airline
flights between airports.  As they point out, the volume of traffic along
each route in this network contains important information about the
operation of the air transport system, but it is not the case that airports
linked by high-volume routes are necessarily close or similar.  In many
cases indeed the reverse is true.  Traffic between Los Angeles and Tokyo is
very high, but this does not mean that Los Angeles and Tokyo are similar
places, or that they are close to one another---they are not.  In this
section therefore, we will consider specifically those networks in which
the weights on edges take greater values for vertex pairs that have closer
connections or are more similar in some way.

The algorithm of Girvan and Newman is based on the idea of betweenness and
works as follows.  The \defn{edge betweenness} of an edge in a network is
defined to be the number of geodesic (i.e.,~shortest) paths between vertex
pairs $s,t$ on the network that run along that edge, summed over all $s$
and~$t$.  If there are two geodesic paths joining a given vertex pair, then
each one counts as a half of a path, and similarly for three or more.  The
edge betweenness is a natural generalization to edges of the well known
vertex betweenness of Freeman~\cite{Freeman77}.  Edge betweenness is high
for edges that act as ``bottlenecks'' for traffic moving about the network.
If traffic from one part of the network to another has to go along one or a
few edges that connect those parts then the betweenness on those edges will
be high.  But this is precisely what we need to find the edges that connect
different communities.  Inter-community edges are precisely those few that
connect otherwise unconnected network portions.  Thus if we remove edges
with high betweenness scores, we will remove the inter-community edges and
leave only the communities themselves behind.

In practice the algorithm is implemented as follows.  We first calculate
the edge betweenness of all edges in the network, using for instance the
fast betweenness algorithm described in~\cite{Newman01c}.  Then we find the
edge that has the highest betweenness and remove it from the network.  If
two or more edges tie for highest betweenness we remove all of them.
Then---and this is crucial---we recalculate the betweenness of all edges on
the remaining network and repeat the process.  As we have argued
elsewhere~\cite{GN02} the recalculation is important for the correct
operation of the algorithm, since it allows for the (common) situation in
which there is more than one edge between a given pair of communities.

How do we generalize this algorithm to the case of weighted networks?
Perhaps the most obvious approach to take would be to generalize the edge
betweenness.  One can define paths on a weighted network by assuming the
``length'' of an edge to vary inversely with its weight, so that two
vertices that are connected twice as strongly will be half as far apart.
Geodesics on such a network can be calculated, for instance, using
Dijkstra's algorithm~\cite{AMO93}.  Then we can define the betweenness of
an edge to be again the number of geodesics between vertices $s,t$ that
pass along that edge summed over all $s$ and~$t$.  And the community
structure algorithm is then one in which we repeatedly remove the edge
having the highest such betweenness and recalculate the betweennesses.

Although an obvious and straightforward generalization of the original
method, however, this algorithm will, almost certainly, give poor results.
To see this, notice that any two vertices that are particularly strongly
connected to one another will have a particularly short distance along the
edge between them.  Geodesic paths will thus, all other things being equal,
prefer to flow along such an edge than along another longer edge between
two less well connected vertices, and hence closely connected pairs will
tend to attract a lot of paths and acquire high betweenness.  This means
that, as a general rule, we are more likely to remove edges between well
connected pairs than we are between poorly connected pairs, and this is the
precise opposite of what we would like the algorithm to do.  Presumably
pairs of vertices that are particularly strongly connected together should
be placed in the same community within the network, but the algorithm as we
have described it deliberately separates such pairs, with the result that
they will often end up in different communities.

Abandoning this approach, therefore, we ask what the correct generalization
is of the algorithm of Newman and Girvan to a weighted network.  To derive
an answer we employ our mapping from the weighted network to a multigraph.
Suppose we have a weighted network with integer weights on the edges and as
before we replace each edge of weight $n$ by $n$ parallel edges of unit
weight.  The adjacency matrix remains unchanged.  Now we apply the normal
unweighted version of our algorithm to the resulting multigraph.

First, we note that the shortest path between any two vertices is
unchanged; since all edges still have unit length any path that was
previously a geodesic is still a geodesic.  However, there are now, in
general, more geodesic paths than there were previously because of the
multiedges.  For each pair of vertices with a double edge running between
them, there are now two geodesics for every one that previously passed
between those vertices---one going along either of the two alternate routes
created by the multiedge.  As before, we count each of these geodesics as a
half of a path.  Thus each of the two edges now has a half of the edge
betweenness that it would have on a simple unweighted graph.  The same
argument applies to multiedges with three or more parallel edges: the
betweenness of each of the parallel edges is equal to the betweenness of
the corresponding edge on the simple graph without multiedges, divided by
the multiplicity of the edge.

Now, following the prescription of the algorithm, we find the edge in the
graph with the highest betweenness and remove it.  But notice that if the
edge removed is a member of a multiedge, then every other member of that
multiedge must have the same betweenness, and hence we should remove all of
them simultaneously.  Thus we end up always removing an entire multiedge at
each step of the algorithm (or more than one if there is a tie for highest
betweenness).  Then, as before, we recompute the betweennesses for all
edges and repeat.

Another and simpler way of summarizing this algorithm is the following: we
calculate the betweennesses of all edges in our weighted graph in the
normal way, ignoring the weights.  Then we divide each such betweenness by
the weight of the corresponding edge, remove the edge with the highest
resulting score, recalculate the betweennesses, and repeat.  We have
derived this algorithm here only for the case of integer weights on the
edges, but we can extend it to the non-integer case using the same trick
that we employed in Section~\ref{maxflow}, of defining successively smaller
units in which the weight of an edge is measured.  The resulting algorithm
is identical to that for the integer case: betweennesses are simply divided
by the weight of the edge and the edge with the highest resulting score
removed from the network.

This algorithm is simple, it is almost as fast as the original unweighted
version (adding only the extra operation of division by the edge weight),
and, as we now show, appears to work excellently.

\begin{figure}
\begin{center}
\resizebox{\figurewidth}{!}{\includegraphics{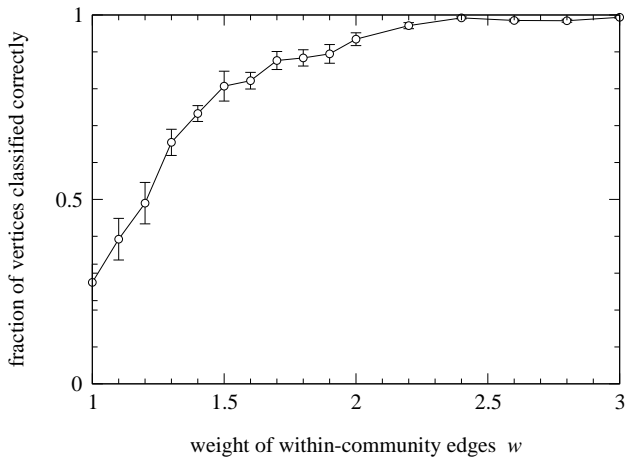}}
\end{center}
\caption{The fraction of vertices classified correctly by our algorithm in
the computer-generated graphs described in the text.  Each point is an
average over ten different graphs.}
\label{rg}
\end{figure}

As a first example of the working of the algorithm we test it on a set of
computer-generated networks.  We have generated random networks of 128
vertices each divided into four groups of 32.  Edges were placed such that
on average each vertex has as many connections to vertices outside its own
group as it does inside.  The mean degree in these tests was fixed at 16.
Unsurprisingly, the normal unweighted community structure algorithm is
unable to pick out communities in networks of this kind, as was
demonstrated previously in Ref.~\cite{GN02}.  Now, however, without
changing the structure of the networks, we assign weights to the edges:
between-community edges are given a fixed weight of~1, while
within-community edges are given a weight~$w\ge1$, which is slowly
increased from a starting value of~1 to explore the sensitivity of the
algorithm.  Fig.~\ref{rg} shows the fraction of vertices classified
correctly by our new algorithm in these tests.

As the figure shows, the result of increasing the within-community
weight~$w$ is immediate: even for very small increases, the algorithm's
performance improves markedly, and more than three quarters of the vertices
are correctly classified for any weight $w>1.5$.  In other words, the extra
information contained in the edge weights does indeed help us enormously to
discern the community structure in the network, and the generalized
algorithm presented here, when given this information, does a good job of
finding that structure.  For values of $w$ greater than 2, the algorithm
classifies essentially all vertices correctly.

Moving to real-world networks, we turn for our second example to a
well-known study from the social networks literature.  In 1972,
Sade~\cite{Sade72} published a network study of a group of sixteen rhesus
monkeys.  Social ties between the monkeys were deduced from grooming
behavior and the study is unusual in that it recorded not only which
monkeys groomed which others, but also the number of instances of grooming
of each monkey by each other during the period of observation.  The result
is a weighted network containing far more information than a simple binary
adjacency matrix.  Grooming forms a directed network between monkeys; one
monkey grooms another and the direction is believed to be associated with
relative status of the individuals.  But for the present study, in which we
regard grooming in either direction as evidence of social interaction, we
have symmetrized the network, creating an undirected one with integer edge
weights equal to the total number of grooming instances in either direction
between each pair of monkeys.  The network has 16 vertices and 69 edges
with edge weights ranging from 1 to~49.

\begin{figure}
\begin{center}
\resizebox{7cm}{!}{\includegraphics{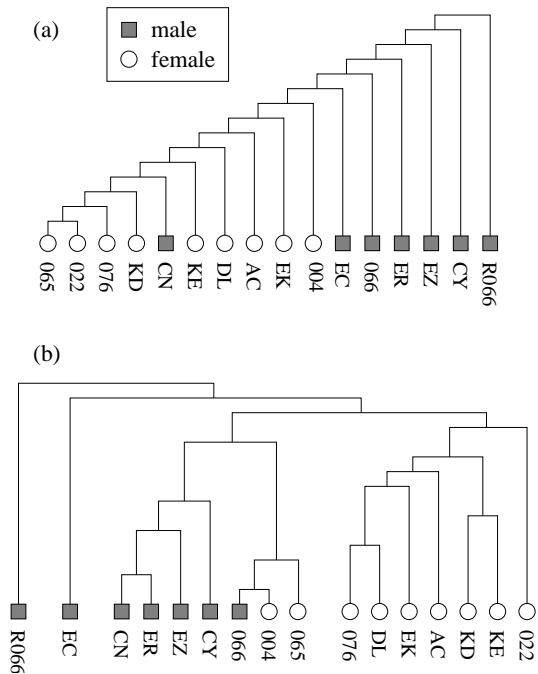}}
\end{center}
\caption{Community structure in the network of sixteen rhesus monkeys
studied by Sade~\cite{Sade72}.  Squares and circles represent male and
female monkeys respectively and the node labels are the same as those used
by the original researcher.  (a)~Dendrogram produced by the algorithm
of~\cite{GN02}, which ignores the weights on the edges.  (b)~Dendrogram for
the algorithm described here, which takes the weights into account.}
\label{rhesus}
\end{figure}

In Fig.~\ref{rhesus}a we show the result of feeding this network through
the ordinary unweighted version of the community structure algorithm, which
takes account only of the presence of edges and not of their weights.  The
results are shown in the form of a tree or ``dendrogram'' of the kind used
in Ref.~\cite{GN02}, which displays the order of the splits in the network
produced by the successive removal of edges.  As the figure shows, the
algorithm finds no community structure at all in the network in this case.
In Fig.~\ref{rhesus}b on the other hand, we show the results of processing
the algorithm through the weighted community structure algorithm, and the
difference is striking.  Now the algorithm detects clear structure within
the group, finding two principal communities, one of females and the other
primarily of males, plus two ``outsider'' males who are not part of either
community.  This accords well with the known social organization of the
monkeys: females tend to associate closely in matrilineal groups; males
tend to associate with one another and with temporary mating partners, but
the adult males also move between tribes every few years (presumably a
tactic to avoid inbreeding within tribes) and outsider males like those
observed here are not uncommon.

These examples suggest that our algorithm is effective at extracting
community structure from weighted networks, including cases in which
algorithms that ignore edge weights find no such structure.  But there is
still a problem: the algorithm does not tell us how many communities a
network should be split into.  The method gives us only a succession of
splits of the network into smaller and smaller communities as represented
by the dendrograms of Fig.~\ref{rhesus}, but it gives no indication of
which splits are best.  In our previous work on unweighted networks, we
solved this problem by introducing a quantity we called the
\defn{modularity}~\cite{NG04}.  This quantity is defined as the fraction of
edges that fall within communities minus the expected value of the same
quantity if edges are assigned at random, conditional on the given
community memberships and the degrees of vertices.

Suppose we have a possible division of an unweighted network into
communities, as provided for example by the algorithm of Ref.~\cite{GN02}.
Let $c_i$ be the community to which vertex~$i$ is assigned.  Then the
fraction of the edges in the graph that fall within communities, i.e.,~that
connect vertices that both lie in the same community, is
\begin{equation}
{\sum_{ij} A_{ij} \delta(c_i,c_j)\over\sum_{ij} A_{ij}}
  = {1\over2m} \sum_{ij} A_{ij} \delta(c_i,c_j),
\end{equation}
where the $\delta$-function $\delta(u,v)$ is 1 if $u=v$ and 0 otherwise,
and $m=\half\sum_{ij} A_{ij}$ is the number of edges in the graph.  If we
preserve the degrees of vertices in our network but otherwise connect
vertices together at random, then the probability of an edge existing
between vertices $i$ and $j$ is $k_ik_j/2m$, where $k_i$ is the degree of
vertex~$i$.  Thus the modularity~$Q$, as defined above, is given by
\begin{equation}
Q = {1\over2m} \sum_{ij} \biggl[ A_{ij} - {k_ik_j\over2m} \biggr]
    \delta(c_i,c_j).
\label{modularity}
\end{equation}
In practice this is an excellent guide to whether a particular division of
a network into communities is a good one.  It takes a value of zero if a
division has no more within-community edges that one would expect by random
chance---a good indication that the division in question is poor.  Nonzero
values indicate deviations from randomness and values around 0.3 or more
usually indicate good divisions.  The maximum possible value of $Q$ is~1.

The same idea can be used to judge community divisions in weighted
networks.  If we apply our rule for mapping weighted networks to
multigraphs, it is straightforward to show that the correct generalization
of the modularity is given by precisely the same formula,
Eq.~\eref{modularity}, provided $A_{ij}$ represents the weight of the edge
between $i$ and~$j$, the degree~$k_i$ is defined according to
Eq.~\eref{degree}, and $m=\half\sum_{ij} A_{ij}$ as before.

The combination of the generalized community structure algorithm and the
generalized modularity allows us now to make definitive divisions of
networks into communities: we apply the algorithm to generate a dendrogram
and then from the divisions represented by the different levels in the
dendrogram we choose the one that gives the highest value of the
modularity.

For a real-world demonstration of this method we take a non-social network,
for a change.  Networks of the co-occurrence of words in bodies of text
have been studied by a number of authors
recently~\cite{FS01a,DM01b,MDLD02}, and are a useful quantitative tool for
analyzing the semantic content of documents.  An influential recent example
of such an analysis is the study by Dooley and Corman~\cite{DC04} of news
reports in the aftermath of the September 11, 2001 attacks in New York and
Washington.  They studied Reuters newswire reports for 66 days following
the attacks and tabulated the occurrence of the commonest words in those
reports by day.  Here we take a typical network from the middle of the
period of the study, the day of Wednesday October 17, 2001.  The vertices
in the network represent words or phrases occurring more than ten times in
wire stories for that day (excluding very commonly occurring words such as
pronouns and prepositions), edges represent the occurrence of pairs of
words in the same sentence, and the weights of the edges are the number of
such occurrences.  The network has a total of 71 vertices and 287 edges,
with edge weights ranging from 1 to 11.  The most commonly co-occurring
pair of vertices is Washington/New York.

\begin{figure}
\begin{center}
\resizebox{\figurewidth}{!}{\includegraphics{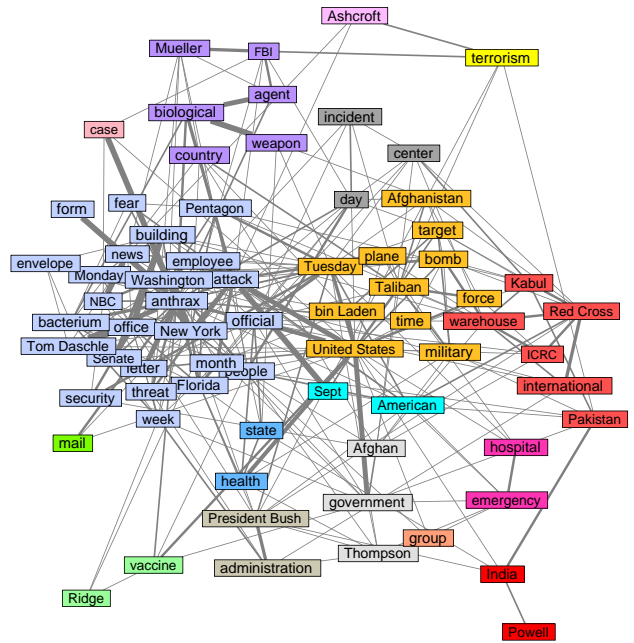}}
\end{center}
\caption{Network of co-occurrence of words in Reuters newswire stories for
October 17, 2001.  The widths of the edges indicate their weights and the
colors of the vertices indicate the communities found by the algorithm
described in the text.}
\label{cra1}
\end{figure}

Making use of these weights in the weighted version of the community
structure algorithm and employing the weighted version of the modularity,
we find that the optimal modularity is achieved for the division into 17
communities shown in Fig.~\ref{cra1}.  The two dominant news stories on
this particular day were the ongoing invasion of Afghanistan by US and
British forces and the anthrax mail attacks taking place in Washington, DC.
As the figure shows, our method clearly picks out these two topics as the
main ``communities'' in the co-occurrence network (left and center-right in
the figure, respectively).  A number of other lesser topics of discussion
are highlighted in the smaller communities: Bush/administration,
Mueller/FBI, international/Red Cross, and so forth.

An analysis of the same network using the unweighted version of the
algorithm finds some of the same structure, but not all of it.  The largest
group of vertices, representing words dealing with the anthrax attacks, is
picked out quite clearly.  The group dealing with Afghanistan is not
however, and the smaller groups make much less sense.  This comes as no
surprise.  Presumably most of the information contained in this network is
in the weights of the edges.  Almost any pair of words might co-occur in a
sentence somewhere in this large body of text, but words that co-occur
frequently---as many as 11 times in this case---almost certainly indicate
linked concepts.

It is worth mentioning that the ideas of this section could easily be
extended to other algorithms for detecting community structure.  Quite a
number of such algorithms have been proposed in recent
years~\cite{Krause03,WH04a,Radicchi04,Newman04a,CSCC04,DM04,OS04}, and in
theory any of these could be generalized to the case of weighted graphs.

\section{Conclusions}
In this paper we have addressed the topic of weighted networks---networks
in which the edges between vertices carry weights representing their
strength or capacity.  Although such networks appear at first to be
substantially more difficult to understand than their unweighted
counterparts, we have argued that in many cases a mapping of the weighted
network onto an unweighted multigraph will allow us to apply directly the
results and techniques developed for the unweighted case.  We have given a
number of examples of this idea, ranging from the very simple, such as
generalizations of degree and eigenvector centrality, to the more complex,
such as the proposal of a new algorithm for detecting community structure
in weighted networks.

The methods presented in this paper are not intended as a rigorous program
for the study of weighted networks, but more as a guide to the intuition
when thinking about these systems.  We look forward with interest to
learning of other applications of these ideas.

\begin{acknowledgments}
The author would like to thank Michelle Girvan and Fred Jin for useful
conversations and Linton Freeman for providing the data for the network of
rhesus monkeys.  This work was supported in part by the National Science
Foundation under grant number DMS--0234188, by the James S. McDonnell
Foundation, and by the Santa Fe Institute.
\end{acknowledgments}

\end{document}